\documentclass{iau} 
\usepackage{graphicx}

\newcommand{\teff}{$T_{\!\mbox{\scriptsize\em eff}}$}

\newcommand{\msun}{$M_\odot$}

\newcommand{\stellarmass} {$M_*$}

\newcommand\aj{AJ}%                                         % Astronomical Journal

\title[Extragalactic Stellar Spectroscopy] %% give here short title %%
{Quantitative Spectroscopy of the Young Stellar Population in Star Forming Galaxies  }

\author[Rolf Kudritzki \& Miguel Urbaneja]   %% give here short author list %%
{Rolf Kudritzki$^{1,2}$
 \and Miguel A. Urbaneja$^3$}

\affiliation{$^1$Institute for Astronomy, University of Hawaii, \\ 2680 Woodlawn Drive,
Honolulu, Hawaii 96822, USA \\ email: {\tt kud@ifa.hawaii.edu} \\[\affilskip]
$^2$ University Observatory Munich, Munich University, \\ Scheiner Str. 1,
81679 Munich, Germany \\[\affilskip]
$^3$Institut f\"ur Astro- und Teilchenphysik, Universit\"at Innsbruck,
    Technikerstr. 25/8, 6020 Innsbruck, Austria}

\pubyear{2018}
\volume{347}  %% insert here IAU Symposium No.
\jname{Early Science with ELTs}
\editors{N. Przybilla, K. Pollard \& A. Calamida, eds.}
\begin{document}

\maketitle

\begin{abstract}
  The determination of chemical composition is crucial for investigating the formation and evolution of star forming galaxies and provides a powerful tool to constrain the effects of galactic winds and accretion from the cosmic web. In this regard stellar absorption line studies provide an attractive alternative to the standard techniques using the strong emission lines of HII regions. We discuss a number of newly developed methods\\
   - multi-object spectroscopy of individual blue and red supergiant stars, the brightest stars in the universe at visual and NIR wavelengths,\\
   - NIR spectroscopy of super star clusters,\\
   - optical spectroscopy of the integrated light of stellar populations in the disks of star forming galaxies,\\
  and present results accumulated over the last two years. We then discuss the scientific perspectives and potential of these methods for the use of ELTs.  
%% add here a maximum of 10 keywords, to be taken form the file <Keywords.txt>
\end{abstract}

\firstsection % if your document starts with a section,
              % remove some space above using this command.
\section{Introduction}

The formation and evolution of galaxies is an extremely complicated process. Gas flows from the intergalactic medium provide the fuel for star formation. Stars through the nuclear fusion processes in their interior produce heavy elements (hereafter ``metals''), which are recycled to the interstellar medium (ISM) by a variety of complex stellar mass-loss processes. While stars continue to form, metals accumulate during the life of a galaxy, but at the same time a significant fraction is also expelled from the ISM by large scale galactic winds. Given the complexity of these many mechanisms and their interplay it is surprising that an intriguingly simple relationship seems to exist between total galactic stellar mass, the mass-metallicity relationship (``MZR''), see for instance \cite{lequeux79}, \cite{tremonti04}, \cite{kud12}. This MZR appears like a Rosetta stone to understand the key aspects of galaxy evolution. For instance, \cite{zahid14} analyzing the evolution of the MZR with cosmological redshift show that the observed MZRs at different redshift can be explained by a model with galactic winds and accretion where the observed metallicity is a straightforward analytical function of the ratio of galactic stellar to ISM gas mass. Galaxies over their lifetime evolve along the (redshift dependent) main sequences of star formation versus stellar mass and turn gas into stars. The low mass metal poor galaxies are gas rich and the high mass metal rich galaxies are gas poor.

\begin{figure}[b]
% \vspace*{-2.0 cm}
\begin{center}
 \includegraphics[width=4.0in]{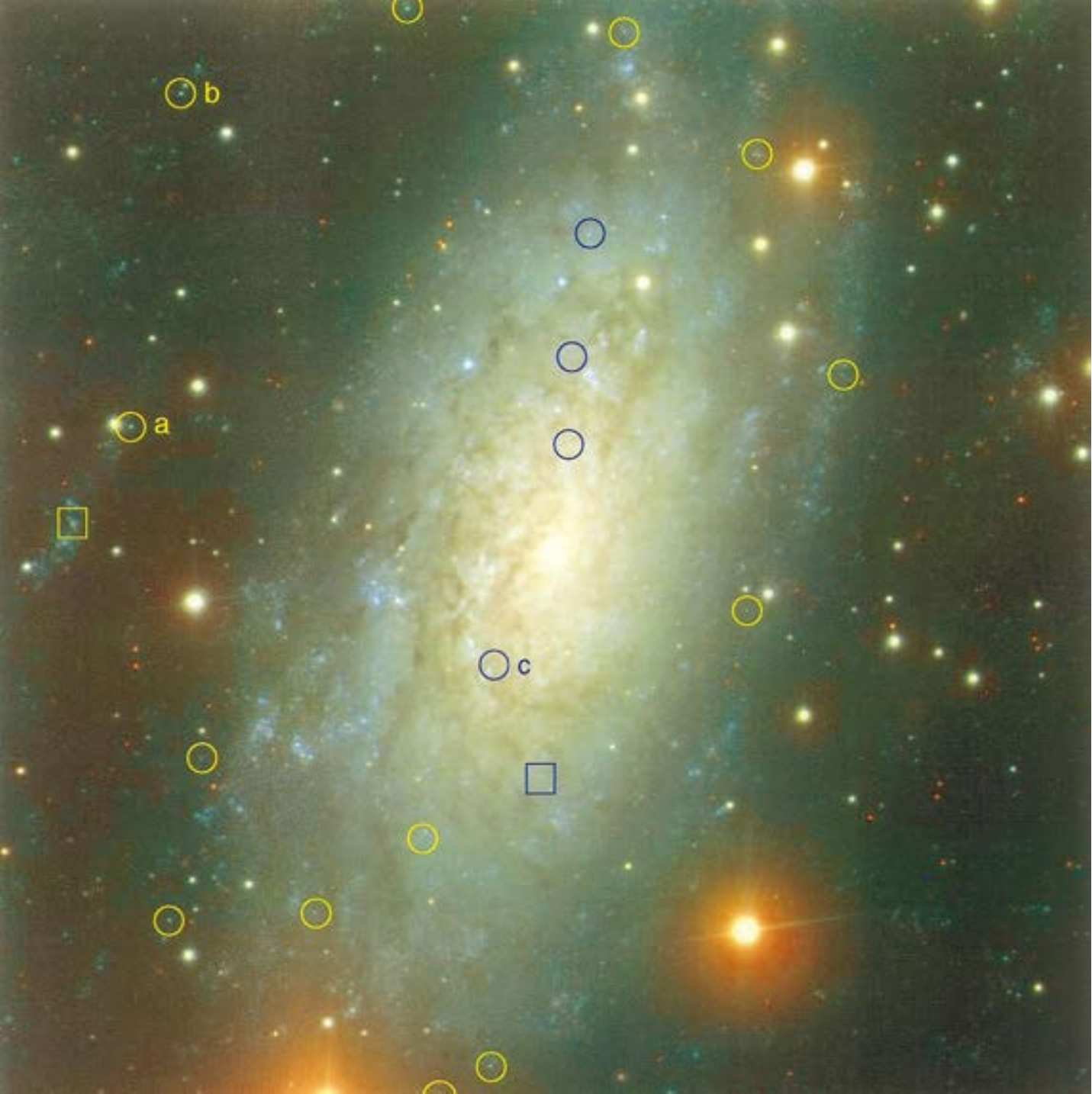} 
% \vspace*{-1.0 cm}
 \caption{
Color composite VLT/FORS image of the spiral galaxy NGC~3621 at 6.5 Mpc distance. The encircled objects are BSG. From \cite{bresolin01}.
}
   \label{fig1}
\end{center}
\end{figure}

Many alternative approaches have been made to use the MZR as an observational constraint of galaxy formation and evolution, \cite{dave11a}, \cite{dave11b}, \cite{yates12}, \cite{dayal13}, \cite{schaye15} to name only a few, and they all seem to be successful. However, there is a fundamental problem. The observed ISM gas-phase oxygen abundances, which are used in these studies to constrain the galaxy evolution models, are affected by enormous systematic uncertainties, which are poorly understood. They are obtained from flux measurements of the strongest ISM emission lines, which are then turned into oxygen abundances using simplifying assumptions of recombination theory. These ``strong line methods'' depend crucially on the calibration methods used and on the sample of HII-regions and galaxies used for the calibration. In a key paper \cite{kewley08} have demonstrated that the systematic uncertainties in oxygen abundances can be as large as 0.8 dex and that the MZR can turn from a steep relationship to an almost flat one for exactly the same data set of emission line fluxes but different calibrations used.

A similar situation is encountered, when the spatially resolved distributions of ISM oxygen abundance of star forming galaxies are investigated. This is crucial step beyond the simple approach just to study global metal abundances of galaxies as a whole and provides important information to constrain the role of galactic winds and accretion. As is well known for decades
(e.g. \cite[Searle 1971]{searle71}; \cite[Garnett \& Shields 1987]{garnett87}; 
\cite[Vila-Costas \& Edmunds 1992]{vilacostas92}; 
\cite[Zaritsky, Kennicut \& Huchra 1994]{zaritsky94}; 
\cite[Garnett, Shields, Skillman et al. 1997]{garnett97};
\cite[Skillman 1998]{skillman98}),
disk galaxies show oxygen abundance gradients with increased metallicities towards the centers and low metallicities in the outskirts. For instance, \cite{ho15} analyzed IFU data of 49 local field star-forming galaxies and found a Gaussian distribution of oxygen abundance gradients with a mean and standard deviation of -0.39$\pm$0.18 dex R$_{25}^{-1}$, where R$_{25}$ is the isophotal radius in the photometric B-band. Applying a statistical distribution of chemical evolution models they concluded that the effects of winds and accretion, while present, must be small. \cite{kud15} extended this work and investigated metallicity and metallicity gradients of 20 disk galaxies in detail to determine the rate of mass-gain and mass-loss through accretion and winds for each of galaxy in the sample. They found three groups of galaxies, one with mostly winds and only very weak accretion, another with mostly accretion and very weak winds and a third group, where winds and accretion were roughly equal. In all three groups the rates of accretion and mass-loss were smaller or at most equal to the star formation rate. However, as in the case of HII-region metallicities from integrated galaxy spectra this result depends crucially on the calibration used for the strong lines. \cite{bresolin09} have demonstrated that spatially resolved HII-region oxygen abundances of disk galaxies and their gradients determined with strong line methods show systematic uncertainties up to 0.6 dex depending on the calibration applied. 

\begin{figure}[b]
% \vspace*{-2.0 cm}
\begin{center}
 \includegraphics[width=4.0in]{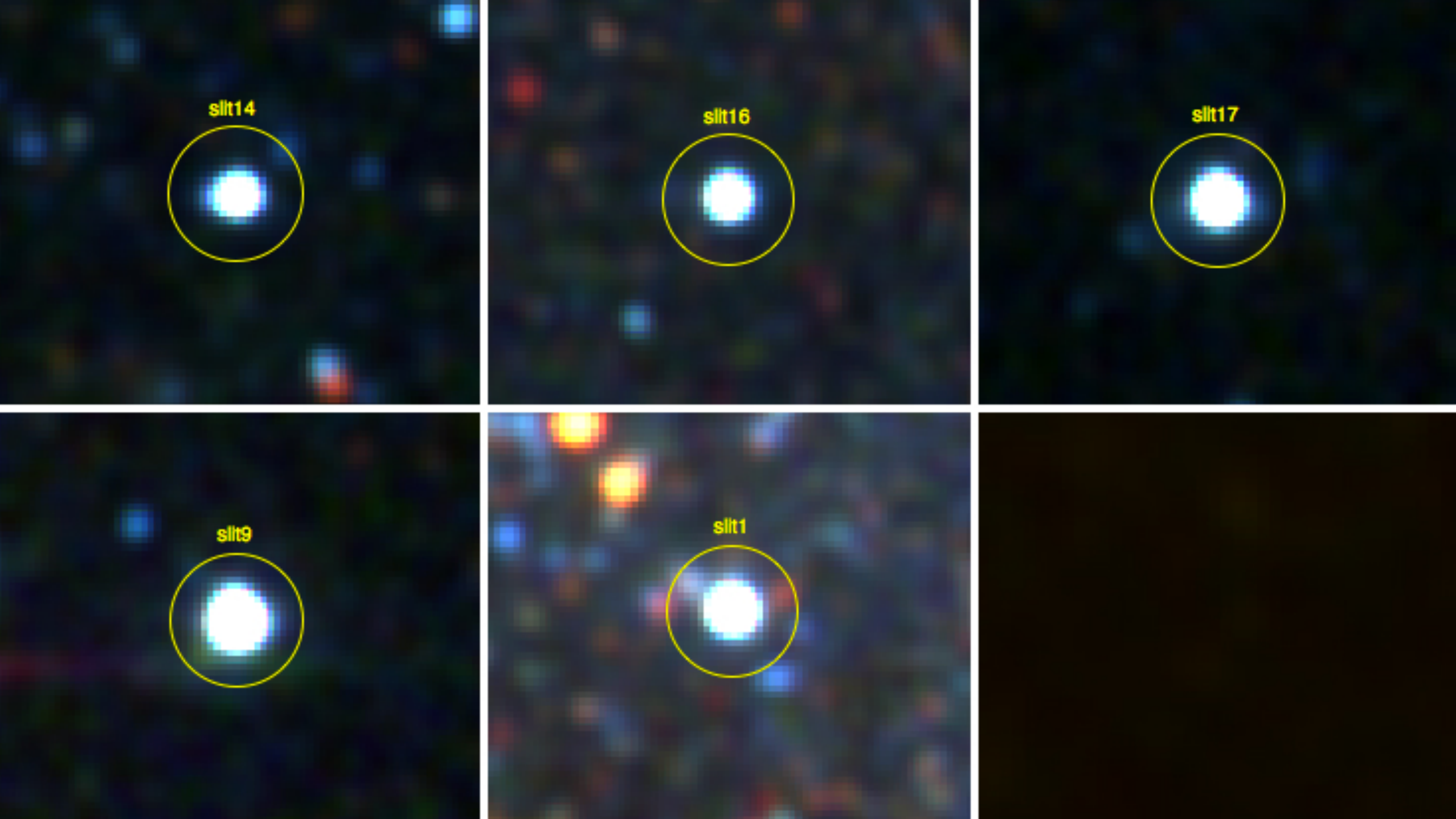} 
% \vspace*{-1.0 cm}
 \caption{Color composite HST ACS images of five BSG in NGC 3621. The circles have a diameter of 1 arcsec. From \cite{kud14}.}
   \label{fig2}
\end{center}
\end{figure}

Of course, an alternative to HII-region strong line methods is the use of weak collisionally excited auroral lines, which allow an estimate of electron temperatures and, thus, eliminate one important source of uncertainty. Unfortunately, because of the weakness of the auroral lines their use is restricted to nearby galaxies or at larger distances to metal poor galaxies. In addition, there is a clear indication of systematic uncertainties at the 0.15 dex level 
(see \cite[Simon-Diaz \& Stasinska 2011]{simondiaz11}; \cite[Zurita \& Bresolin 2012]{zurita12}; 
\cite[Gazak, Kudritzki, Evans et al. 2015]{gazak15}; 
\cite[Bresolin, Kudritzki, Urbaneja et al. 2016]{bresolin16}).

Thus, in view of the crucial role of accurate metallicity determinations for the investigation of the formation and evolution of galaxies the development of alternative methods is of fundamental importance. The obvious alternative is to use the information contained in the absorption line spectra of stars. It has long been the dream of stellar astronomers to quantitatively analyze the spectra of individual stellar objects in distant galaxies and to determine stellar parameters, distances, metallicity and interstellar reddening and extinction. With the advent of 8-10m class telescopes and the efficient optical and NIR multi-object spectrographs attached to them this dream has become true and with the new future generation of ELTs we it will turn into the important new era of extragalactic stellar astronomy.

\begin{figure}[b]
% \vspace*{-2.0 cm}
\begin{center}
 \includegraphics[width=4.0in]{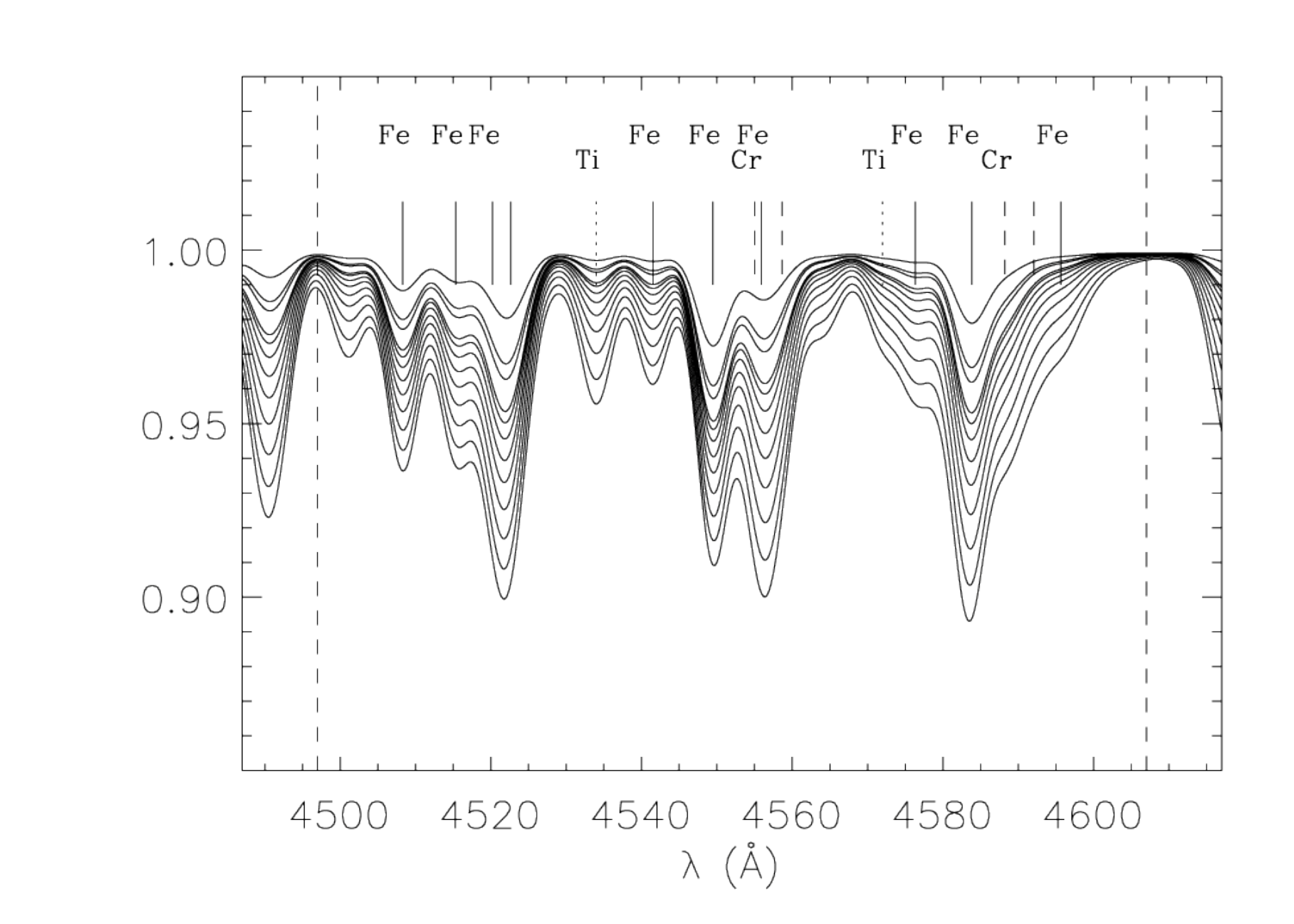} 
% \vspace*{-1.0 cm}
 \caption{
Synthetic metal line spectra of a typical A supergiant calculated as a function of metallicity in the spectral window from 4497~\AA~to 4607~\AA. Metallicities range from [Z] = -1.30 to 0.30. The synthetic spectra were folded with a Gaussian of 5~\AA~FWHM to 
account for the instrumental broadening. Rotational broadening is negligible. From \cite{kud08}.}
   \label{fig3}
\end{center}
\end{figure}

The most promising objects for such studies are massive stars in the mass range of 15 to 40 \msun in the evolutionary phase, when they leave the main sequence and cross the HRD as blue supergiants of spectral B and A and then become red supergiants of spectral type K and M. In the next three sections, we describe the results of quantitative extragalactic supergiant spectroscopy accomplished over the last decade. We then introduce a new and very promising method how to use the absorption lines of the integrated stellar spectra in disk galaxies for a determination of the metallicity of the young stellar population and we conclude with a brief outlook into the ELT era.

\begin{figure}
\begin{center}
 \includegraphics[width=4.0in]{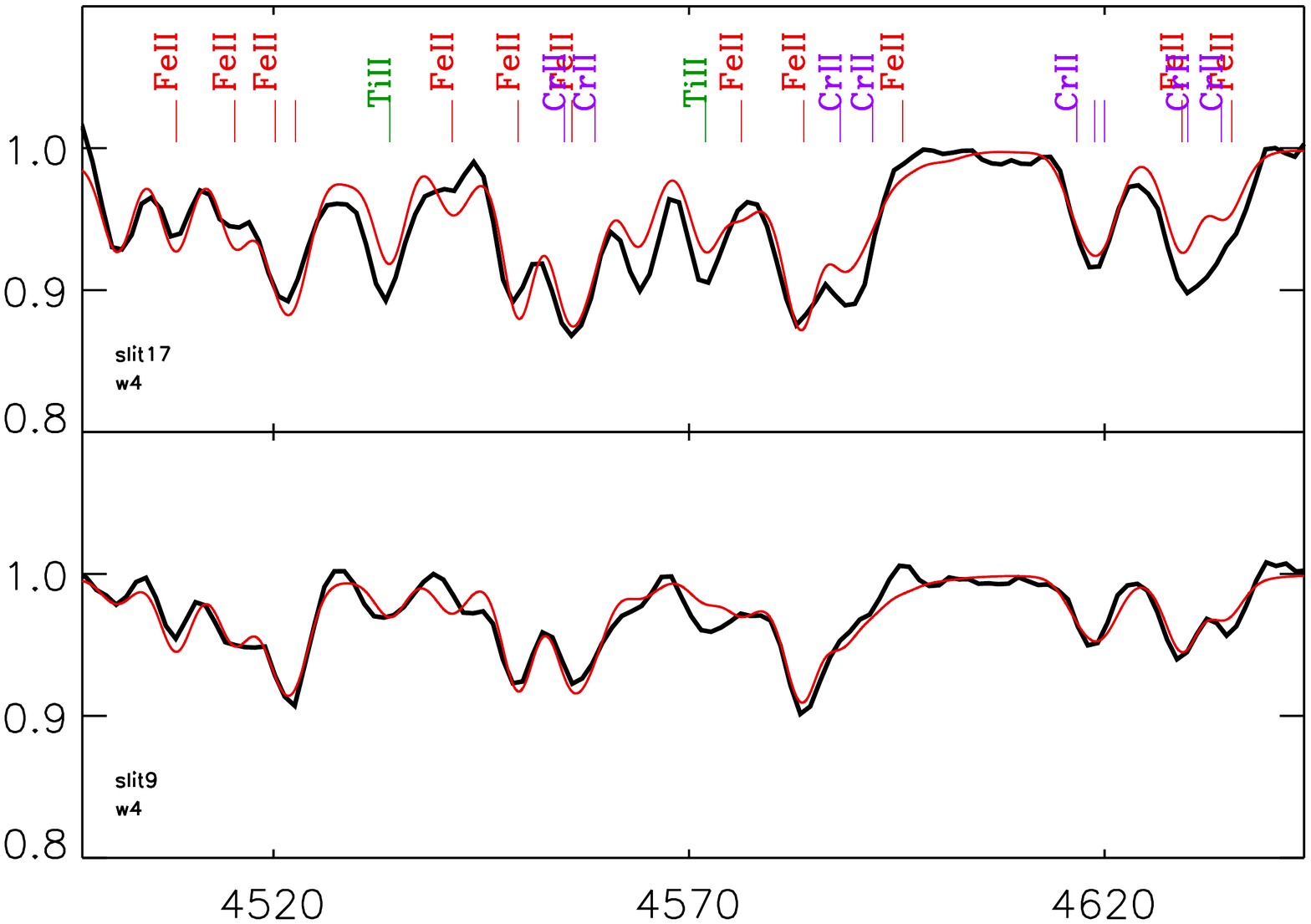}
 \includegraphics[width=4.0in]{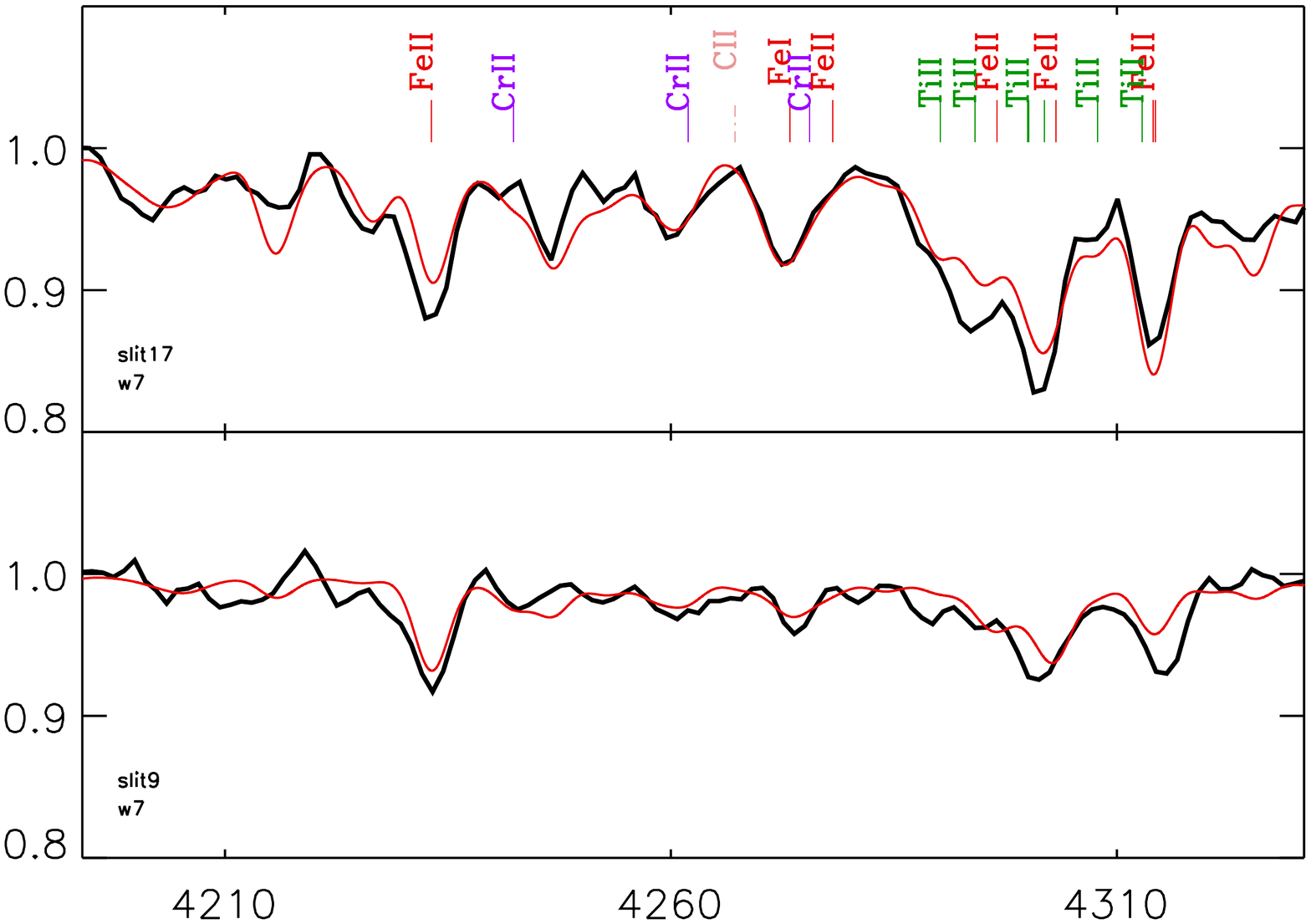}
\caption{Spectral fit of two A supergiant BSG in NGC~3621 (observed
spectra in black, model spectra in red). Only two spectral windows are shown. The analysis includes the full availabe spectrum from 3800 to 5500~\AA. From \cite{kud14}.}  \label{fig4}
 \end{center}
\end{figure}

\section{Blue Supergiant Stars}

When massive stars evolve through the HRD at almost constant luminosity towards the red supergiant stage, there visual brightness reaches a maximum at spectral types B and A because of the variation of bolometric correction with effective temperature as a result of Wien's displacement law. In consequence, these blue supergiant stars (BSG) are the optically brightest stars in the universe. They can reach absolute magnitudes up to $M_{V} \cong -10$, which means that one single star can become as bright as the integrated light of a globular cluster or a dwarf galaxy. In a star forming galaxy they can easily be detected (see Fig.\,\ref{fig1}) and they shine out everything else in their surroundings (Fig.\,\ref{fig2}), which makes them ideal for quantitative spectroscopic studies even in the relatively dense fields of galactic disks. With an age between 10 to 50 Myr their metallicity represents the one of the ISM out of which they just formed.

\begin{figure}[b]
% \vspace*{-2.0 cm}
\begin{center}
 \includegraphics[width=4.0in]{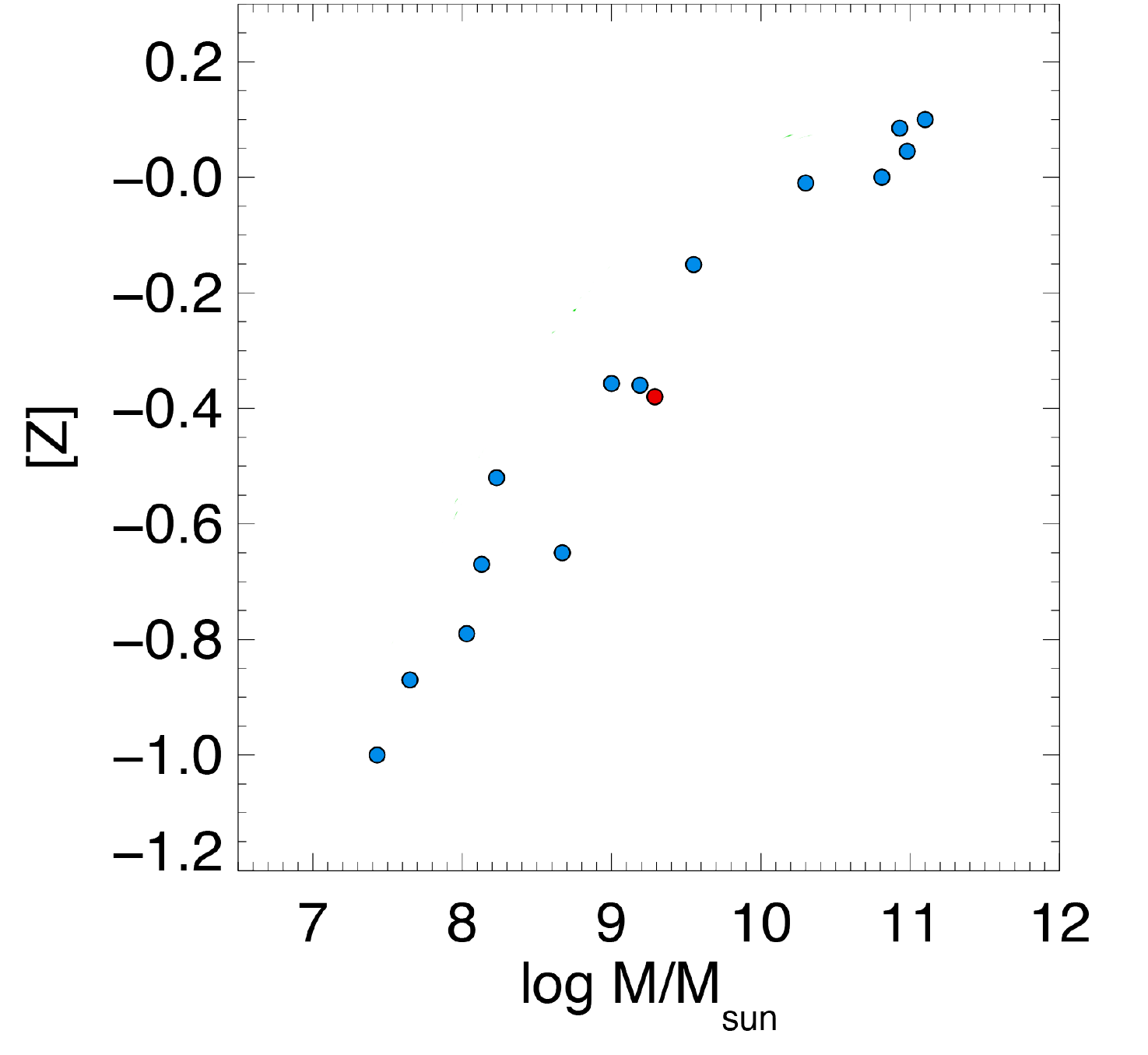} 
% \vspace*{-1.0 cm}
 \caption{
The relationship between galactic stellar mass and metallicity, MZR, as obtained from quantitative spectroscopy of supergiant stars. From \cite{kud16}.}
   \label{fig5}
\end{center}
\end{figure}

The spectral analysis of BSG is not easy. The extremely low gravity of their photospheres requires an appropriate inclusion of non-LTE effects as first shown by \cite{kud73}. However, with a considerable effort involving hundreds of men years of hard work by many colleagues mostly at University Observatory Munich the framework for the accurate non-LTE spectroscopy analysis of hot massive stars has been developed. A good example is the work by \cite{przybilla06}, who analyzed high resolution, high S/N spectra of Milky Way A supergiant stars using very detailed NLTE line formation calculations including ten thousands of lines in NLTE (see also 
\cite[Przybilla, Butler, Kudritzki et al. 2000]{przybilla00}; 
\cite[Przybilla, Butler, Becker et al. 2001]{przybilla01a}; 
\cite[Przybilla, Butler \& Kudritki 2001]{przybilla01b};
\cite[Przybilla \& Butler 2001]{przybilla01};
\cite[Przybilla 2002]{przybilla02}) to determine stellar parameters and
abundances with hitherto unkown precision ($\Delta$\teff\/ to $\le 2$\%, log~g to $\sim 0.05$ dex,
individual metal abundances to $\sim 0.05$ dex). The study by \cite{urbaneja17} gives an impression of similar work at lower metallicity. 

The next step towards the extragalactic application is the modification of the BSG analysis technique for spectra with low resolution. BSG are extremely bright, but at distances clearly beyond the Local Group it the S/N obtainable with high resolution is insufficient for quantitatitive spectroscopic work. Luckily, even at the low resolution of 5 $\AA$ of multi-object spectrographs such as FORS at the VLT or LRIS at Keck accurate spectroscopic analysis is still possible. Fig.\,\ref{fig3} shows model spectra degraded to this low resolution illustrating how the changes of the strength of crucial metal lines with metallicity are still clearly observable provided the S/N is high enough. The low resolution analyis technique for BSG of spectral types A and late B has been developed by \cite{kud08}, \cite{kud13} and \cite{hosek14}. In parallel, \cite{urbaneja05} developed the method for BSG of early B spectral type. Fig.\,\ref{fig4} shows a typical fit of metal lines in a spectrum of a BSG in a galaxy at 6.5 Mpc distance resulting in a metallicty accurate to 0.1 dex.

Using these techniques a considerable number of galaxies has now been studied: NGC\,300 -- \cite{kud08}; WLM -- \cite{bresolin06}; 
\cite{urbaneja08}; NGC\,3109 -- \cite{evans07}, \cite{hosek14}; IC\,1613 --
\cite{bresolin07}, \cite{berger18}; NGC\,6822 -- \cite{muschielok99}, \cite{przybilla02}; M33 -- \cite{u09}; NGC\,55 -- \cite{castro12}, \cite{kud16}; M81 -- \cite{kud12}; NGC\,4258 -- \cite{kud13}; M83 -- \cite{bresolin16}. The results allowed for a detailed discussion of stellar evolution, chemical evolution and the determination of accurate distances using the flux-weighted gravity-- luminosity relationship (FGLR) developed and calibrated by \cite{kud03}, \cite{kud08} and \cite{urbaneja17}. Most importanly, they have also been used to construct the first MZR of galaxies based solely on stellar spectroscopy (Fig.\,\ref{fig5}). This new and very accurate MZR can now be used as a constraint for galaxy evolution modeling and to re-calibrate HII-region strong line methods for galaxy metallicity measurements.

\begin{figure}[b]
% \vspace*{-2.0 cm}
\begin{center}
 \includegraphics[width=4.0in]{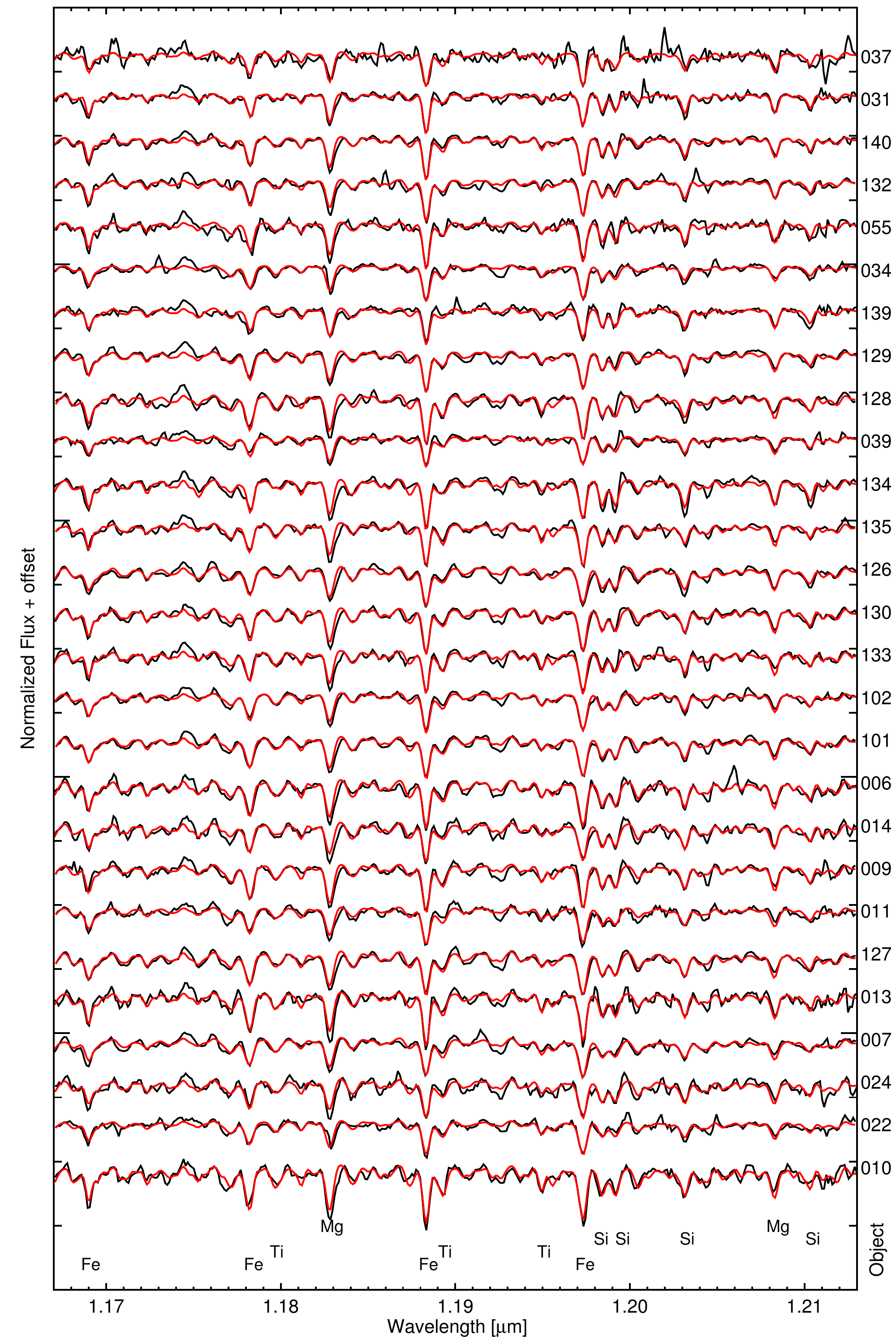} 
% \vspace*{-1.0 cm}
 \caption{
KMOS J-band spectra of RSG in the Sculptor spiral galaxy NGC~300 (black) and non-LTE spectral fits (red). Key atomic metal lines are indicated at the bottom. From \cite{gazak15}.}
   \label{fig6}
\end{center}
\end{figure}

\begin{figure}[b]
% \vspace*{-2.0 cm}
\begin{center}
 \includegraphics[width=4.0in]{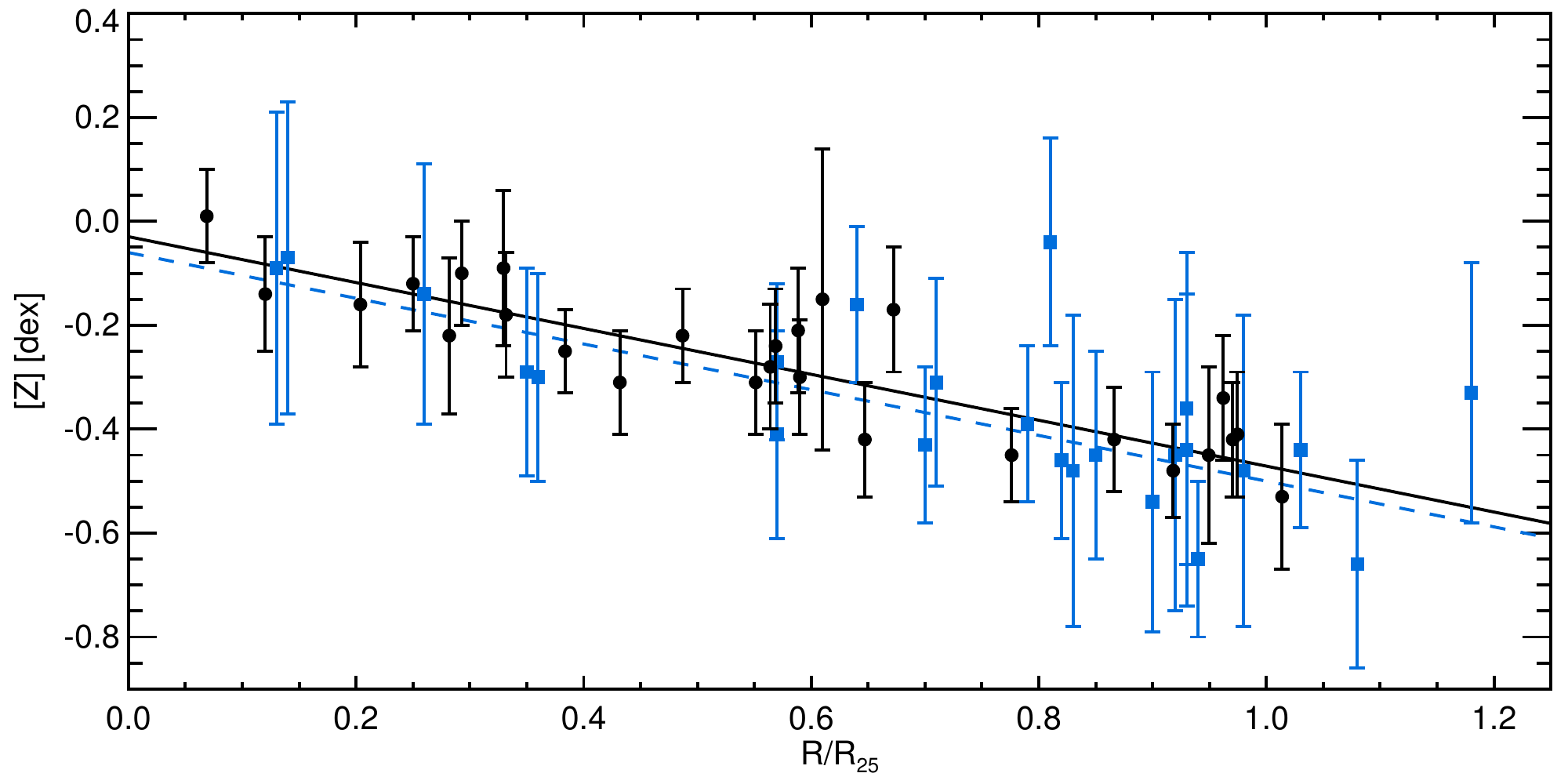} 
% \vspace*{-1.0 cm}
 \caption{
Metallicity versus galacto-centric distance of supergiant stars in NGC~300. R$_{25}$ is the B-band isophotal radius of NGC~300 and R is the galactro-centic distance after de-projection. Blue: BSG from the optical analysis of VLT/FORS spectra by \cite{kud08}, black: RSG form VLT/KMOS J-band analysis by \cite{gazak15}. The dashed blue and solid black regressions show the the metallicity gradients obtained from BSG and RSG, respectively. From \cite{gazak15}.}
   \label{fig7}
\end{center}
\end{figure}

\section{Red Supergiant Stars}

Red supergiant stars (RSG) are the immediate successors of BSGs in massive star evolution and pile up at the HRD Hayashi-limit of fully convective stars with high luminosities at very low effective temperatures. With absolute magnitudes up to $M_{J} \cong -11$ they are the brightest stars in the universe at NIR light. Belonging to the same stellar population as BSG they provide an excellent alternative for extragalactic metallicity studies with spectrographs such as MOSFIRE at Keck or KMOS at the VLT. However, the challenge is to develop an accurate analysis technique at intermediate or low spectral resolution for these cool objects, where the spectra are densely crowded with atomic and molecular lines.

\cite{davies10} investigated this problem and found that the spectral analysis sweet spot is the J-band at a resolution of R $\sim$ 3000. The  J-band is dominated by isolated atomic metal lines an and relatively void of molecular lines. Moreover, as shown in \cite{davies13}, the J-band is least affected by photospheric surface inhomogeneities caused by the giant convective cells of RSGs. As demonstrated in the hydrodynamic models by \cite{chiavassa11} the effects caused by these inhomogeneities are devastating for modelling optical line spectra with strong lines but are moderate in the NIR. \cite{davies13} confirmed this conclusion. They found that effective temperatures derived from strong optical lines where in severe disagreement with observed energy distributions, while results obtained from fitting J-band lines agreed reasonably well.

The spectroscopic J-band technique has been tested carefully with high resolution, high signal to noise spectra of RSG in Per OB1 by \cite{gazak14b} and in the LMC and SMC by \cite{davies15}. The metallicities determined agreed well with independent spectroscopic studies of BSG or young main sequence stars. Detailed tests degrading the spectral resolution down to the MOSFIRE/KMOS value of 3000 confirmed that the method continues to be accurate also at intermediate resolution. The line formation calculations used for the analysis include a detailed treatment of non-LTE effects based on the work by \cite{bergemann12}, \cite{bergemann13}, and \cite{bergemann15}.

The first step beyond the Local Group in the quantitative spectral analysis of RSG applying the J-band technique has been carried out by \cite{gazak15} in the Sculptor galaxy NGC~300. Fig.\,\ref{fig6} gives an impression of the spectra and the quality of the model fits. Fig.\,\ref{fig7} compares the results obtained for individual RSG and BSG. As can be seen from the regressions, the results obtained from RSG and BSG are almost identical.

Fig.\,\ref{fig7} is a triumph of quantitative extragalactic stellar spectroscopy. The analysis methods of BSG and RSG are entirely different. BSG are hot stars with optical spectra dominated by ionic lines and non-LTE radiative transfer models adapted to hot stars. RSG are cools stars with NIR spectra dominated by atomic lines and the non-LTE models adapted to cools stars. Yet, the results are in perfect quantitative agreement. This gives us confidence that the systematic errors inherent in the stellar spectral analysis of supergiant stars must be small. On the other hand, the comparison of the BSG and RSG results with HII-region oxygen abundances based on collisonally excited weak auroral lines as carried out by \cite{bresolin09} reveals a systematic effect with 0.1 dex lower metallicity. This is very likely an effect of oxygen depletion into HII-region dust.

New comprehensive J-band work on RSGs in the galaxies NGC~6288, NGC~55 and the LMC cluster NGC~2100 by \cite{patrick15}, \cite{patrick16}, \cite{patrick17} confirms the close agreement between the RSG and BSG spectroscopic techniques.

\begin{figure}[b]
% \vspace*{-2.0 cm}
\begin{center}
 \includegraphics[width=4.0in]{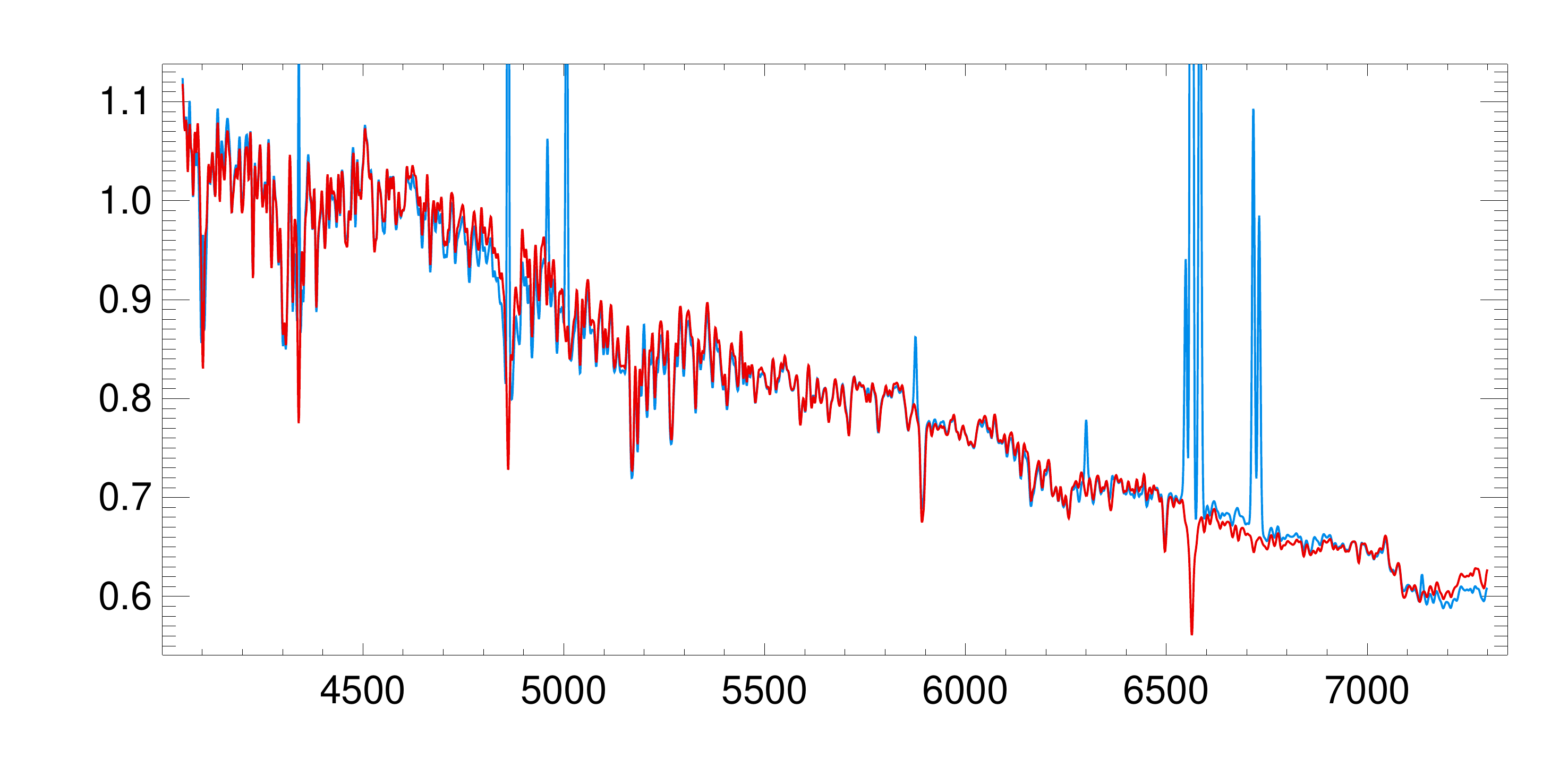} 
% \vspace*{-1.0 cm}
 \caption{
Fit of the stacked spectrum of SDSS star forming galaxies with stellar mass log\stellarmass~= 10.0 (blue) with a sequential burst population synthesis model (red). Only the stellar metal lines are considered for the fit, the spectral regions with ISM emission lines are not included. See \cite{zahid17}.}
   \label{fig8}
\end{center}
\end{figure}

\begin{figure}[b]
% \vspace*{-2.0 cm}
\begin{center}
 \includegraphics[width=4.0in]{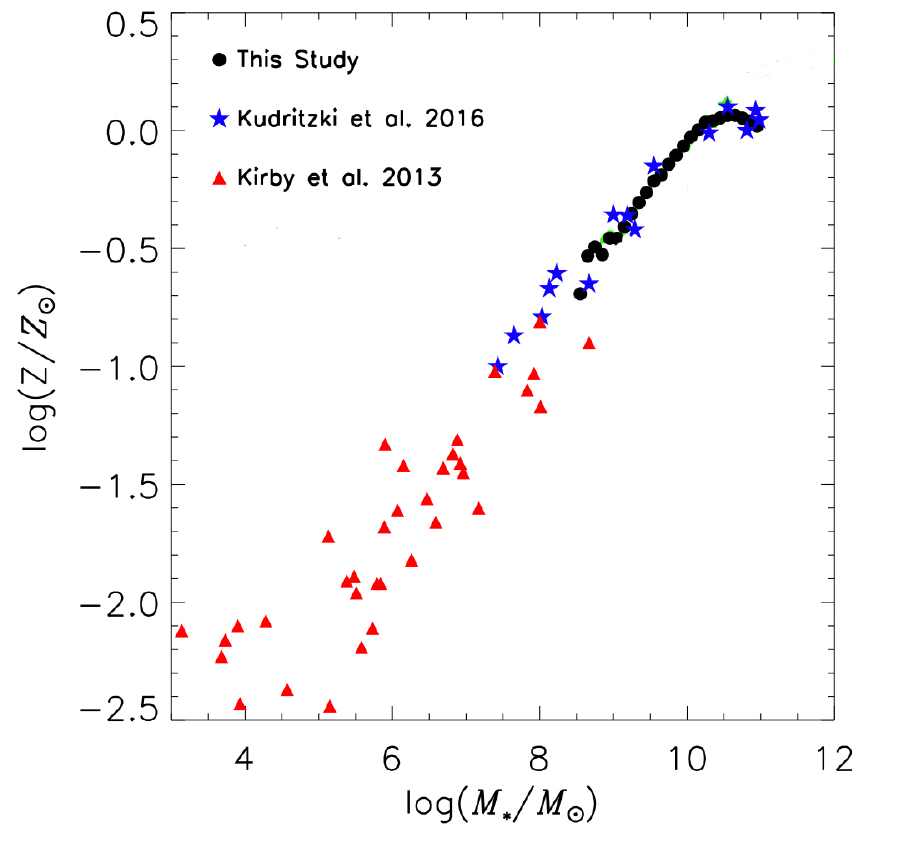} 
% \vspace*{-1.0 cm}
 \caption{
The MZR of SDSS star forming galaxies obtained from sequential burst population synthesis (black) compared to spectroscopy of individual supergiant stars (blue stars) and giant stars in Local Group dwarf galaxies (red triangles). Modified figure from \cite{zahid17}.}
   \label{fig9}
\end{center}
\end{figure}

\section{Red Supergiants in Super Star Clusters}

Super Star Cluster (SSC) are coeval young and compact agglomerates of stars with total stellar masses in excess of 10$^4$ up to 10$^6$~\msun (\cite{portegies10}), which are mostly found in galaxies with high star formation rates. In a pioneering study \cite{gazak13} and \cite{gazak14a} have shown that, as soon as SSCs are older than 7 Myr, their NIR light is dominated by massive red supergiant stars. For instance, for a cluster with 10$^5$ \msun~of the order of 100 RSG exist at any time and provide more than 95\% of the J-band light. Since RSG span only a narrow range in effectice temperature, the J-band spectrum of an SSC older than 7 Myr looks like a spectrum of a single red supergiant. SSC are, thus, ideal targets for extragalactic metallicity determinations in the same way as RSG but hundred times brighter.

\cite{gazak14a} have pioneered this technique with a J-band metallicity study of two SSC, one in the metal rich galaxy M83 and one in the metal poor spiral NGC~6946. \cite{gazak14b} and \cite{patrick16} have tested the method by combining observed J-band spectra of individual RSG in the massive galacitic cluster Per OB1 and the LMC cluster NGC~2100, respectively, into a single spectrum simulating an unresolved SSC and analyzing it with respect to metallicity. They found excellent agreement with the metallicity derived from the individual RSG, thus, confirming the simple concept to analyze SSC J-band spectra by using model calculations for single RSG. \cite{davies17} carried out the ultimate test of the SSC method by investigating a large sample of SSC in M83 and by comparing with metallicities obtained from the optical analysis of single BSG. The close agreement found confirmed that SSC through J-band spectroscopy are excellent tracers of metallicity. They also used their new results on M83 to compare mass-metallicity relationships obtained from RSG and SSC with those from BSG and constructed a new MZR based on the spectroscopy of massive stars. \cite{lardo15} studied clusters in the Antennae galaxy NGC 4038 at 20 Mpc distance, for the first time applying the J-band techniqe at large distances. 

\section{Integrated Spectra of Star Forming Galaxies}

The unresolved, spatially integrated spectra of star forming galaxies contain two spectral components, the strong emission lines from the ionized ISM gas and a stellar continuum with absorption lines. While the former have been intensively used for metallicity studies, as discussed in the introduction, only a few attempts have been made to investigate the latter, see for instance \cite{cid14}. Recently, \cite{zahid17} carried out a quantitative absorption line study of 200,000 SDSS star forming galaxies. They stacked the spectra into stellar mass bins log~\stellarmass/\msun~from 8.5 to 11.0 and applied a population synthesis technique, where model spectra of sequential star formation bursts were fitted to the observations to provide the metallicity of young stellar population (Fig.\,\ref{fig8}). They obtained a MZR consistent with the spectroscopy of individual supergiant stars discussed in the previous sections (Fig.\,\ref{fig9}). This technique is extremely promising and is presently extended to investigate the spatial distribution of metallicity in star forming galaxies utilizing the wealth of information available from IFU spectroscopy. Such investigations could go out to large distances, since the the surface brightnesses of galaxies do not change with distance.

\section{Future work with ELTs}

The potential of extragalactic stellar spectroscopy is enormous. Future ELTs will use optical MOS spectrographs supported by GAO and NIR MOS units with MCAO. This pushes the frontier for detailed evolution studies with BSG spectroscopy to 40 Mpc. Even more impressive is the gain for the J-band spectroscopy of RSG and SSC with the help of MCAO. Here, the study by \cite{evans11} indicates that with two nights at an ELT we can investigate individual RSG at 100 Mpc, the distance of the Coma cluster, and SSC could be studied at even ten times larger distances. At the same time IFUs attached to ELTs would allow to investigate the spatial distribution of chemical composition with unprecedented accuracy. These combined methods will open up a huge volume of the local universe for a detailed and accurate investigation of the evolution of galaxies.

\begin{discussion}

\discuss{Miriam Pena}{Have the ISM emission lines of the ionized gas been studied in your spectra? How does the chemical composition compare with the ones of the stars?}

\discuss{Rolf Kudritzki}{For all our spectroscopic stellar studies, individual supergiant stars or integrated spectra, we have always compared with HII-region studies. As for the strong line studies, many of them are off and only very few agree. If the HII-region work includes weak collisonally excited auroral lines, then the agreement is reasonable, however, clear systematic effects of the order of 0.1 dex are encountered. See, for instance, Bresolin, Kudritzki et al., 2016, ApJ 830, 64}

\end{discussion}

\end{document}